\documentclass[twocolumn,epsf,psfig,superscriptaddress]{revtex4}
\usepackage{amssymb}
\usepackage{epsfig}
\DeclareMathAlphabet{\pazocal}{OMS}{zplm}{m}{n}            
\usepackage{graphicx}
\usepackage{color}
\usepackage{bm}
\usepackage[normalem]{ulem}     
\usepackage{soul}
\usepackage{amsmath}
\usepackage{braket} 
\usepackage{rotating}
\usepackage{multirow} 
\DeclareMathAlphabet{\pazocal}{OMS}{zplm}{m}{n}            
\usepackage{graphicx}
\usepackage{hyperref}

\begin{document}
\title{Universal responses in nonmagnetic polar metals}

\author{Fabian J\"ager} 
\affiliation{Materials Theory, ETH Zurich, Wolfgang-Pauli-Strasse 27, 8093 Zurich, Switzerland} 

\author{Nicola A. Spaldin}
\affiliation{Materials Theory, ETH Zurich, Wolfgang-Pauli-Strasse 27, 8093 Zurich, Switzerland}

\author{Sayantika Bhowal} 
\affiliation{Materials Theory, ETH Zurich, Wolfgang-Pauli-Strasse 27, 8093 Zurich, Switzerland} 

\date{\today}

\begin{abstract}
We demonstrate that two phenomena, the kinetic magneto-electric effect and the non-linear Hall effect,
are universal to polar metals, as a consequence of their coexisting and contraindicated polarization
and metallicity. We show that measurement of the effects provides a complete characterization of the nature
of the polar metal, in that the non-zero response components indicate the direction of the polar axis, and the
coefficients change sign on polarization reversal and become zero in the non-polar phase.
We illustrate our findings for the case of electron-doped PbTiO$_3$ using a combination of density functional
theory and model Hamiltonian-based calculations. Our model Hamiltonian analysis provides crucial insight into
the microscopic origin of the effects, showing that they originate from inversion-symmetry-breaking-induced
inter-orbital hoppings, which cause an asymmetric charge density quantified by odd-parity charge multipoles.
Our work both heightens the relevance of the kinetic magneto-electric and non-linear Hall effects, and
broadens the platform for investigating and detecting odd-parity charge multipoles in metals.
\end{abstract}

\maketitle

\section{Introduction}
The idea of combining electric polarization with metallicity, against the common belief that polarization is screened by itinerant carriers, was first conceived by Anderson and Blount more than fifty years ago \cite{AndersonBlount1965}. It has come to reality, however, rather recently with the practical material realization of polar metals \cite{Shi2013, Fei2018, Sharma2019}. 
These have consequently opened up a new paradigm for investigating numerous intriguing physical effects that result from the coexistence of the seemingly mutually exclusive properties of polarity and metallicity \cite{Zhou2020,BhowalSpaldin2023,Daniel2023}.

In the present work, we point out two such effects, the kinetic magneto-electric effect (KME) and non-linear Hall effect (NHE), which are universal to all polar metals. While these effects have been sporadically investigated in some candidate polar metal systems \cite{Ma2019,Kang2019,Xiao2020}, a consensus in applying these effects to characterizing polar metals is still missing.   
Here we show that both these effects carry simultaneously the key signatures of the polar metal phase, that is the direction of the polar axis, the switchability of the polarization, and the ferroelectric-like nonpolar to polar structural transition, and so provide a complete characterization of polar metals. 
Furthermore, we reveal the microscopic origin of these two effects by analyzing asymmetries in the charge density. While both effects are dominated by contributions from the electric dipole moment i.e., the first-order asymmetry in the charge density, the electric octupole moment, characterizing the third-order asymmetry in the charge density, also plays an important role.

The kinetic magneto-electric effect is a linear effect, describing electric field ($\cal E$) induced magnetization ${\cal M}_j = {\cal K}_{ij} {\cal E}_i$ in a nonmagnetic metal \cite{Levitov1984,Yoda2015,Zhong2016,Sahin2018,Tsirkin2018}. The resulting magnetization, in turn, gives rise to a transverse Hall current ($J$) as a second-order response to the applied electric field, $J_i =\chi_{ijk} {\cal E}_j {\cal E}_k$, known as nonlinear Hall effect \cite{SodemannFu2015}. ${\cal K}_{ij}$ and $\chi_{ijk}$ are the KME response and non-linear Hall conductivity (NHC) tensor respectively with $i,j,k$ indicating the Cartesian directions. Within the relaxation-time approximation for the nonequilibrium electron distribution, both these responses can be elegantly recast in terms of the equilibrium  reciprocal-space magnetic (spin plus orbital) moment $\vec m (\vec k)$ and the Berry curvature dipole (BCD) ${\cal D}_{ij}$ respectively \cite{Zhong2016,Tsirkin2018,SodemannFu2015}:
\begin{eqnarray} \nonumber \label{formula}
&&{\cal K}_{ij} = -\frac{e\tau}{\hbar}  \sum_n \frac{1}{(2\pi)^3} \int d^3k~ m^n_j (\vec k) \partial_{k_i} \epsilon^n_k \Big( \frac{\partial f_0}{\partial \epsilon^n_k }\Big) 
= \frac{e\tau}{\hbar} \tilde{\cal K}_{ij} \\ \nonumber
&&  \chi_{ijk} =-\varepsilon_{ilk} \frac{e^3\tau}{2(1+i\omega\tau)} {\cal D}_{jl} \\ 
&&= \varepsilon_{ilk} \frac{e^3\tau}{2(1+i\omega\tau)} \sum_n \frac{1}{(2\pi)^3}\int d^3k~ \Omega^n_l (\vec k) \partial_{k_j} \epsilon_k \Big( \frac{\partial f_0}{\partial \epsilon^n_k }\Big). 
\end{eqnarray}    
Here $\tau$, $\varepsilon_{adc}$, $e, n$, $f_0$ and $\vec \Omega$ are respectively the relaxation time-constant, Levi-Civita symbol, the electronic charge, band index, equilibrium Fermi distribution function and Berry curvature. Both the reduced KME response $\tilde{\cal K}_{ij}$ and the BCD are intrinsic properties of a material and are given by,\cite{Zhong2016,Tsirkin2018,Ma2019,Kang2019}
\begin{eqnarray} \label{response1} \nonumber 
\tilde{\cal K}_{ij} &=& -\sum_n \frac{1}{(2\pi)^3} \int d^3k~ m^n_j (\vec k) \partial_{k_i} \epsilon^n_k \Big( \frac{\partial f_0}{\partial \epsilon^n_k }\Big)\\ 
&=& \sum_n \frac{1}{(2\pi)^3}\int d^3k~ (\partial_{k_i} m^n_j) f_0 
\end{eqnarray}
and, 
\begin{eqnarray} \label{response2}\nonumber
{\cal D}_{ij} &=& -\sum_n \frac{1}{(2\pi)^3}\int d^3k~ \Omega^n_j (\vec k) \partial_{k_i} \epsilon_k \Big( \frac{\partial f_0}{\partial \epsilon^n_k }\Big) \\ 
&=& \sum_n \frac{1}{(2\pi)^3}\int d^3k~ (\partial_{k_i} \Omega^n_j) f_0 .
\end{eqnarray} 
Both $\tilde{\cal K}_{ij}$ and ${\cal D}_{ij}$ are allowed in nonmagnetic metals with gyrotropic point group symmetry \cite{Zhong2016,Tsirkin2018,Ma2019,Kang2019}. Since all polar point groups are gyrotropic \cite{gyrotropic,HeLaw2020}, both KME and NHE are allowed by symmetry in all polar metals. 

Interestingly, the components of the reciprocal-space magnetic moment $\vec m (\vec k)$, that contributes to the KME response, is determined by the direction of the electric polarization \cite{Bhowal2022}. Similarly, the antisymmetric component of the BCD, ${\cal D}^- = ({\cal D}-{\cal D}^T)/2$, correlates with the orientation of the polar axis $\vec d$, $d_i \equiv \varepsilon_{ijk} {\cal D}_{jk}^{-}/2$ \cite{SodemannFu2015}, suggesting a possible switching of both responses for a switchable orientation of the polar distortion. Furthermore, since both effects are forbidden by symmetry in an inversion symmetric structure, a structural transition from a centrosymmetric to a noncentrosymmetric polar structure can be inferred from the onset of these effects as the temperature is lowered.  

We illustrate these concepts by explicitly considering the case of electron-doped PbTiO$_3$ (PTO) as an example material. Undoped PTO is a prototypical conventional ferroelectric insulator \cite{Nelmes1985}. Interestingly, even upon electron doping via replacing the Ti$^{4+}$ ions by Nb$^{5+}$ ions, the resulting PbTi$_{1-x}$Nb$_x$O$_3$ was observed to sustain the electric polarization up to $x = 0.12$, at which point the system also becomes conducting \cite{Gu2017,Takashi2000}. 
In the present work, using both first-principles density functional theory (DFT) and a model Hamiltonian-based approach we show that the presence as well as the orientation of the polar axis in the polar metal phase of doped PTO can be determined from the non-zero components of KME and the NHE.  

The remainder of this manuscript is organized as follows. We start by describing the computational details in section \ref{method}. This is followed by the results and discussions in section \ref{results}, where we present our computational results for doped PTO, describing the existence and tuning of KME and NHE, the momentum space distribution of the orbital moment and Berry curvature that determine these effects, their microscopic origin within the model Hamiltonian framework, and the role of odd-parity charge multipoles. Finally, we summarize our results in section \ref{summary} and give a proposal for measuring these effects.   

\section{Computational details}\label{method}

The responses $\tilde{\cal K}_{ij}$ and ${\cal D}_{ij}$ are computed using the QUANTUM ESPRESSO\cite{Giannozzi2009} and Wannier90 codes \cite{MarzariVanderbilt1997,Souza2001,Mostofi2014}. We use fully relativistic norm-conserving pseudo-potentials for all the atoms with the following valence electron configurations: Pb ($6s^26p^2$), Ti ($4s^23d^2$), and O ($2s^22p^4$). Self-consistency is achieved with a 12$\times$ 12$\times$10 $k$-point mesh and a convergence threshold of $10^{-7}$ Ry. 
The {\it ab-initio} wave functions, thus obtained, are then projected to maximally localized Wannier functions \cite{MarzariVanderbilt1997,Souza2001} using the Wannier90 code
\cite{Mostofi2014}. In the disentanglement process, as initial projections, we  choose 42 Wannier functions per unit cell which
include the $s$ and $p$ orbitals of Pb, $d$ orbitals of Ti and $s$ and $p$ orbitals of O atoms, excluding the rest. After the disentanglement is achieved, the wannierisation process is converged to $10^{-10}$ \AA$^2$. We then compute the $k$-space distribution of the orbital moment and the Berry curvature as well as the reduced KME response, $\tilde{\cal K}_{ij}$ and the BCD ${\cal D}_{ij}$ for a 150$\times$ 150$\times$140 $k$-point mesh. To estimate the doped charge density, we also compute the densities of states (DOS) for the same $k$-point mesh.

\section{Results and Discussion} \label{results}

\subsection{NHE and KME and their tuning in polar metals } 

\begin{figure}[t]
\includegraphics[width=\columnwidth]{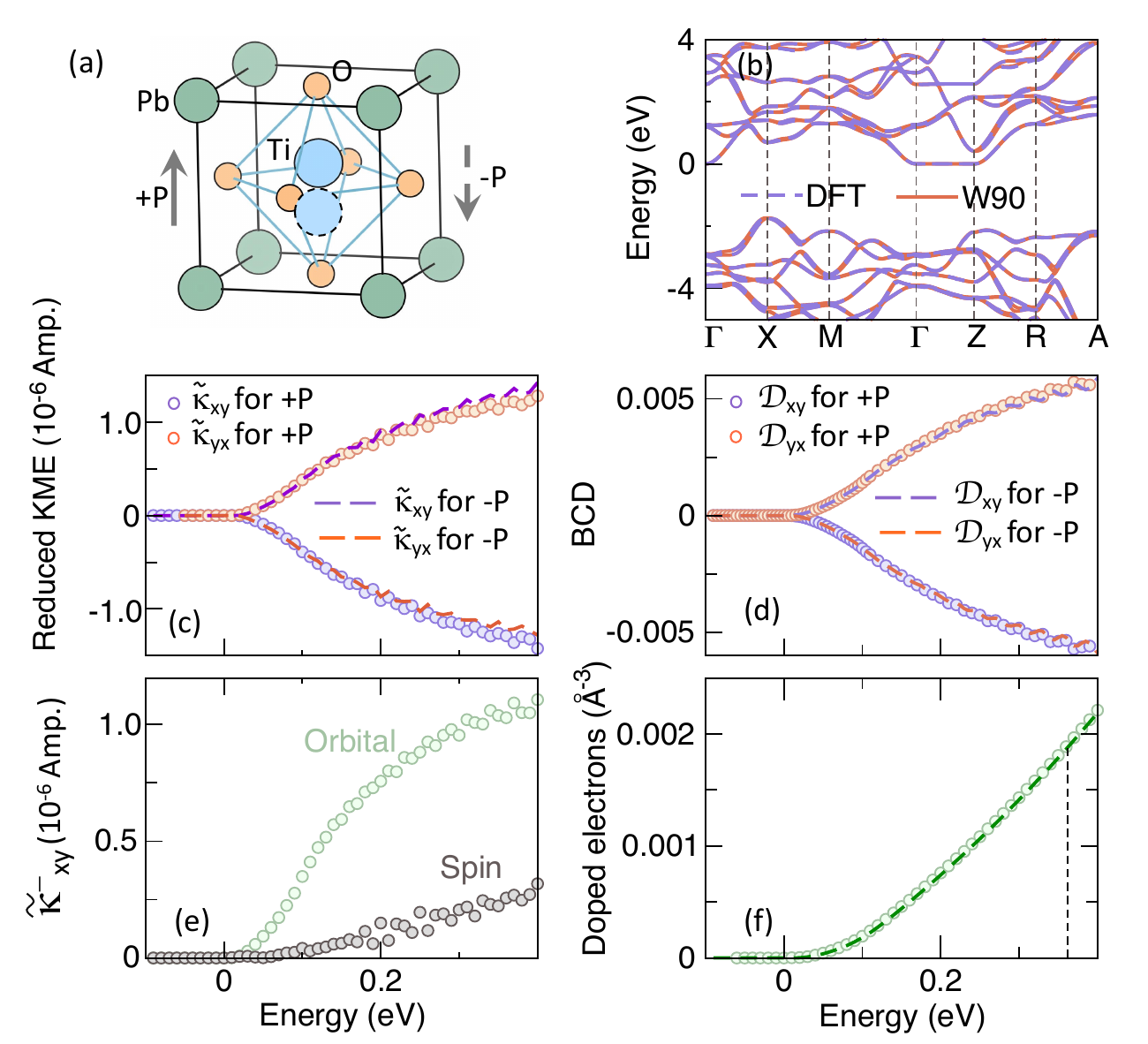}
\caption{(a) Schematic illustration of the crystal structure of PTO, showing the off-centering of the Ti atom leading to a polarization $+P$ along $\hat z$. The dashed circle indicates the displacement of the Ti atom in the opposite direction, switching the direction of the polarization ($-P$) indicated by the dashed arrow. (b) Comparison of the band structure of undoped PTO, computed within DFT (dashed line) and that obtained from Wannier90 (solid line), showing a good agreement between the two. (c) Computed reduced KME response components $\tilde{\cal K}_{xy}$ (blue) and $\tilde{\cal K}_{yx}$ (red) for the two directions of polarization $+P$ (circle) and $-P$ (dashed line), shown in (a), as a function of energy. (d) BCD components ${\cal D}_{xy}$ (blue) and ${\cal D}_{yx}$ (red) for the polarization directions, $+P$ (circles) and $-P$ (dashed line) as a function of energy. (e)  Energy variation of the spin and orbital contributions to the absolute value of the antisymmetric component of the reduced KME response, $\tilde{\cal K}^{-}_{xy}=\frac{1}{2}(\tilde{\cal K}_{xy}-\tilde{\cal K}_{yx}$) for polarization $+P$. (f) Energy variation of the doped electron densities for the two polarization directions, $+P$ (circles) and $-P$ (dashed line). The vertical black dashed line corresponds to the experimentally achieved maximum doped electron density, that maintains the polarity of the structure. The zero of energy in (b)-(f) refers to the CBM of undoped PTO. }
\label{fig1}
\end{figure}

We start with the electronic structure of PTO, which crystallizes in the non-centrosymmetric tetragonal ($P4mm$) structure with the polar $C_{4v}$ point group symmetry \cite{Nelmes1985}. In tetragonal PTO, both Pb$^{2+}$ ($6s^2$ lone pair) and Ti$^{4+}$ ($3d^0$) ions off-center with respect to the surrounding O$^{2-}$ ions, resulting in a net polarization along $\hat z$, which is switchable to $-\hat z$ using an external electric field. We refer to these two structures, schematically depicted in Fig. \ref{fig1} (a), as $+P$ and $-P$ respectively. The electronic structure of the polar undoped PTO (corresponding to $+P$) is shown in Fig. \ref{fig1} (b), depicting the insulating band structure in which the occupied O-$p$ states and the fromally empty Ti-$t_{2g}$ states form the valence band maximum (VBM) and conduction band minimum (CBM) respectively. 

The doping electrons in doped PTO occupy the CBM, leading to a metallic band structure within the rigid band approximation. In order to compute $\tilde{\cal K}_{ij}$ and ${\cal D}_{ij}$, we first project the computed {\it ab-initio} wave functions onto maximally localized Wannier functions, and then disentangle the relevant bands (see section \ref{method} for computational details) from the rest using the Wannier90 code \cite{Mostofi2014}. As depicted in Fig. \ref{fig1} (b), the wannierised bands agree well with the full DFT band structure. The central quantities $\tilde{\cal K}_{ij}$ and ${\cal D}_{ij}$ in determining the magnitudes of the KME and NHE are then computed using Eqs. (\ref{response1}) and (\ref{response2}) as implemented within the Wannier90 code \cite{MarzariVanderbilt1997,Souza2001,Mostofi2014}. 

 \begin{figure}[t]
\includegraphics[width=\columnwidth]{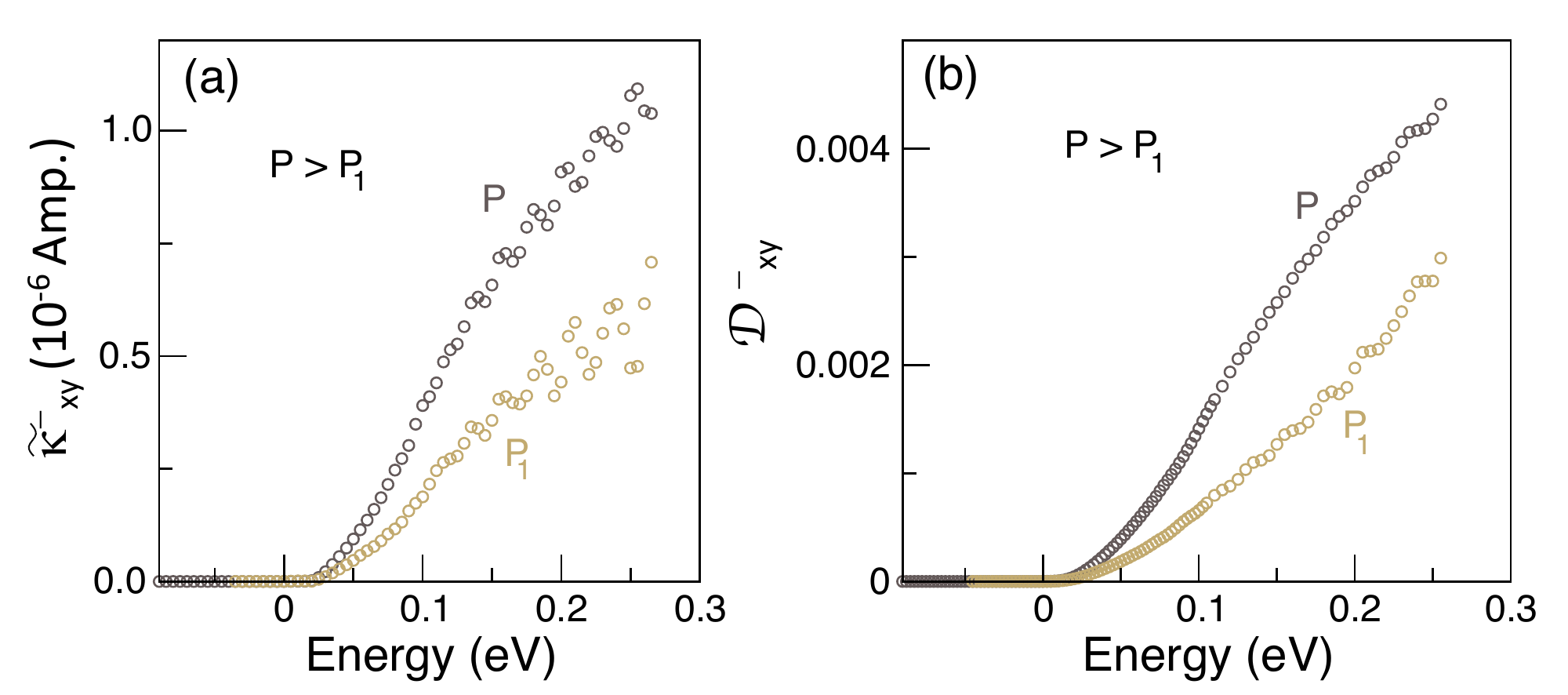}
\caption{(a) Comparison of the energy variation of the absolute value of $\tilde{\cal K}^{-}_{xy}$ for two different displacements of the Ti ion, $P$ and $P_1$, with the former being larger than the latter. (b) Comparison of the energy variation of the absolute value of the antisymmetric BCD component ${\cal D}^{-}_{xy}=\frac{1}{2}({\cal D}_{xy}-{\cal D}_{yx}$) for the same $P$ and $P_1$. }
\label{fig2new}
\end{figure}

The computed non-zero components of the reduced KME response, $\tilde{\cal K}_{xy}$ (blue circle), $\tilde{\cal K}_{yx}$ (red circle), and BCD ${\cal D}_{xy}$ (blue circle) and ${\cal D}_{yx}$ (red circle) are shown as a function of energy in Fig. \ref{fig1} (c) and (d) for the $+P$ structure. To determine whether the energy range used in the computation is experimentally achievable, we further compute the doped electron density by integrating the corresponding DOS and show the results in Fig. \ref{fig1} (f). Note that the zero of the energy corresponds to the CBM for the undoped case. The vertical dashed line in Fig. \ref{fig1} (f) indicates the maximum doped electron density up to which the polarity of the lattice persists in the experiments \cite{Gu2017,Takashi2000}, justifying the chosen energy range.

We note from Fig. \ref{fig1} (c) and (d) that $\tilde{\cal K}_{xy}=-\tilde{\cal K}_{yx}$ and ${\cal D}_{xy}=-{\cal D}_{yx}$, consistent with the $C_{4v}$ point group symmetry. Here $\tilde{\cal K}_{ij}$ has both spin and orbital contributions. In order to understand the relative contributions of the two, the individual spin and orbital contributions are also shown in Fig. \ref{fig1} (e) for the absolute value of the antisymmetric component of the reduced KME response, $\tilde{\cal K}^{-}_{xy}=\frac{1}{2}(\tilde{\cal K}_{xy}-\tilde{\cal K}_{yx})$. This clearly shows that the orbital contribution dominates over the spin contribution. Such a current-induced orbital magnetization has also been reported for other systems with broken inversion symmetry \cite{Yoda2015,He2020,BhowalSatpathy2020,Hara2020} and may have important implications in the field of orbitronics. 

To see the effect of the polarization direction, we reverse the direction of the displacement of the ions, leading to the $-P$ structure (see Fig. \ref{fig1} (a)). The corresponding computed $\tilde{\cal K}_{xy}$ (blue dashed line), $\tilde{\cal K}_{yx}$ (red dashed line), and ${\cal D}_{xy}$ (blue dashed line), ${\cal D}_{yx}$ (red dashed line) are shown in Figs. \ref{fig1} (c) and (d) respectively. We note that in this case, all the computed quantities switch sign compared to the $+P$ structure, while still maintaining the symmetry of the $C_{4v}$ point group, as discussed above. 

We further artificially decrease the amount of the Ti displacement to see the effect of the magnitude of polarization. We refer to the corresponding structure as $+P_1$. The computed absolute values of $\tilde{\cal K}^{-}_{xy}$ and ${\cal D}^{-}_{xy}=\frac{1}{2}({\cal D}_{xy}-{\cal D}_{yx})$ for $+P_1$ are depicted in Figs. \ref{fig2new} (a) and (b) respectively, together with the values for $+P$. We find that both $\tilde{\cal K}^{-}_{xy}$ and ${\cal D}^{-}_{xy}$ have smaller magnitude for $+P_1$ compared to $+P$, suggesting that both effects not only depend on the direction of polarization but also depend on the magnitude of the polarization. 

It is important to point out here that polarization is not the only factor that contributes to the value of the responses. For example, both responses also depend on the details of the electronic structure (see Eqs. \ref{response1} and \ref{response2}). As a result, the situation can be more complicated if there is a drastic change in the band structure with the change in electric polarization. Nevertheless, our analysis clearly shows that the polarization is an important factor and that both KME and NHE are tunable by changing the direction or magnitude of the electric polarization.

 \begin{figure}[b]
\includegraphics[width=\columnwidth]{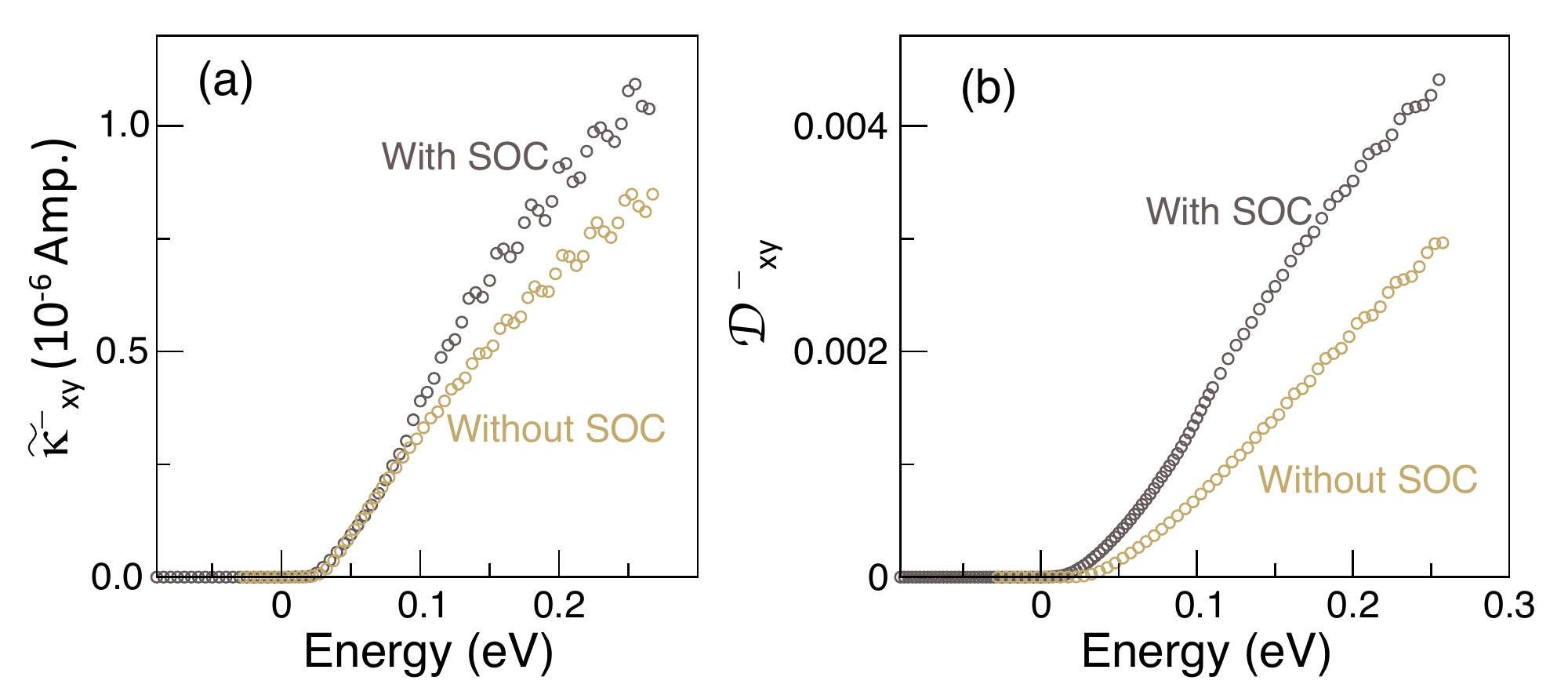}
\caption{Comparison of the energy variation of the absolute value of (a) $\tilde{\cal K}^{-}_{xy}$ and (b) ${\cal D}^{-}_{xy}$ in the absence and presence of SOC. }
\label{fig2}
\end{figure}
%
Furthermore, to understand the dependence on the spin-orbit coupling (SOC), we also perform additional calculations with the SOC turned off in our computations. Comparisons of the computed $\tilde{\cal K}^{-}_{xy}$ and ${\cal D}^{-}_{xy}$ both in the absence and presence of SOC are shown in Figs. \ref{fig2} (a) and (b). As seen from these figures, both $\tilde{\cal K}^{-}_{xy}$ and ${\cal D}^{-}_{xy}$ exist even without the SOC. This suggests that both effects occur due to the symmetry of the structure and the presence of SOC is not necessary. Indeed, in the absence of SOC, the KME response is driven by the orbital contribution. With the inclusion of SOC, the orbital degrees of freedom couple to the spin degrees of freedom, and consequently, it leads to additional current-induced spin magnetization in the system. The inclusion of SOC, therefore, increases the magnitudes of both effects. It is important to point out here that unlike these responses, the spin-splitting of the bands and the resulting unconventional magnetic Compton scattering \cite{BhowalCollinsSpaldin2022} occur only in the presence of SOC.

 \subsection{$k$-space distribution of orbital moment and Berry curvature} \label{distribution}
 \begin{figure}[t]
\includegraphics[width=\columnwidth]{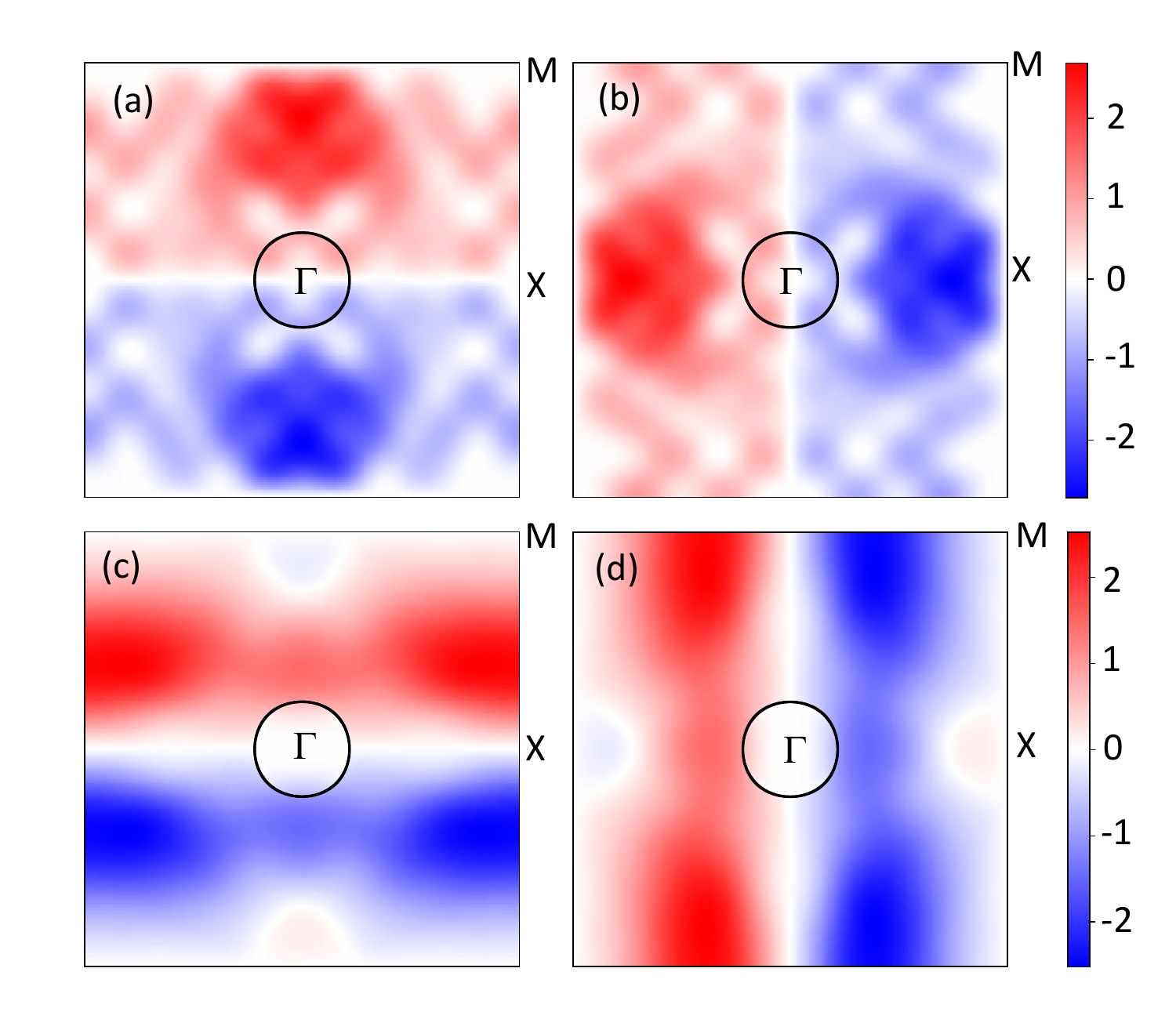}
\caption{Computed $k$-space distributions of the orbital moment components (in units of eV-\AA$^2$) (a) $m^{\rm orb}_x$, (b) $m^{\rm orb}_y$, and the components of the Berry curvature (in units of \AA$^2$) (c) $\Omega_x$, and (d) $\Omega_y$ in PTO for the doped electron density $\sim 0.001$\AA$^{-3}$. The black circle around the $\Gamma$ point indicates the Fermi surface corresponding to this electron density. $\vec {m}^{\rm orb} (\vec k)$ ($\vec \Omega (\vec k)$) is the sum of $\vec m^{n, \rm orb} (\vec k)$ ($\Omega^n_k (\vec k)$), given in Eq. (\ref{morb}) (Eq. (\ref{BC})), over the occupied bands. The high-symmetry $k$-points are indicated on the $k_z = 0$ plane for easy reference.}
\label{fig3}
\end{figure}

To better understand the responses, we further compute the $k$-space distributions of the relevant orbital magnetic moment components $m^{\rm orb}_x(\vec k), m^{\rm orb}_y(\vec k)$ and Berry curvature components $\Omega_x(\vec k), \Omega_y(\vec k)$ in the $k_x$-$k_y$ plane. Since, the $\tilde{\cal K}_{ij}$ response is dominated by the orbital contribution, here for simplicity we only consider the orbital magnetic moment distribution. The orbital magnetic moment is computed within the modern theory by evaluating the expectation value of the orbital magnetization operator $\frac{-e}{2} (\vec r \times \vec v)$ \cite{Xiao2005,Thonhauser2005,Ceresoli2006,Shi2007} with $-e < 0$, as implemented in the Wannier90 code \cite{Lopez2012},
\begin{eqnarray} \label{morb}\nonumber
 \vec m^{n, \rm orb} (\vec k) &=& \frac{e}{2\hbar} {\rm Im} \langle \nabla_k u^n_k|\times [{\cal H}(\vec k)-\epsilon^n_k]|\nabla_k u^n_k\rangle \\ 
 &+& \frac{e}{\hbar} {\rm Im} \langle \nabla_k u^n_k|\times [\epsilon^n_k-\epsilon_{\rm F}]|\nabla_k u^n_k\rangle.
\end{eqnarray}
Here, $\epsilon^n_k$ and $u^n_k$ are the energy eigenvalues and eigenfunctions of the Hamiltonian ${\cal H}(\vec k)$ obtained from Wannierization, 
and $\epsilon_{\rm F}$ is the Fermi energy. We note that since the KME response is a Fermi surface property (see Eq. \ref{formula}), the second term in Eq. \ref{morb} does not contribute to the KME response. This can be seen easily by recognizing that $\frac{\partial f_0}{\partial \epsilon^n_k } = -\delta(\epsilon^n_k-\epsilon_{\rm F})$, and so has a non-zero value only if $\epsilon^n_k=\epsilon_{\rm F}$, in which case the second term in Eq. (\ref{morb}) vanishes.
The $k$-space distribution of the Berry curvature is computed using the Kubo formula \cite{TKNN1982},
\begin{eqnarray} \label{BC}
 \Omega^n_k (\vec k) &=& -2\hbar^2\sum_{m \ne n} {\rm Im} \frac{\langle u^n_k|v_i|u^m_k\rangle \langle u^m_k|v_j|u^n_k\rangle}{(\epsilon_k^n-\epsilon_k^m)^2},
\end{eqnarray} 
where $\vec v = \frac{1}{\hbar} \frac{\partial {\cal H}}{\partial \vec k}$ is the velocity operator and ($i,j,k$) are cyclic permutations of the Cartesian directions ($x,y,z$).

Both $\vec m^{\rm orb} (\vec k)$ and $\vec \Omega (\vec k)$ follow the same symmetry relations: Under spatial inversion $\cal I$ symmetry both remain invariant, with $\vec m^{\rm orb} (\vec k) \xrightarrow{\cal I} \vec m^{\rm orb} (-\vec k)$, whereas under time-reversal ($\cal T$) symmetry they switch sign, $\vec m^{\rm orb} (\vec k) \xrightarrow{\cal T} -\vec m^{\rm orb} (-\vec k)$ (similarly for $\vec \Omega (\vec k)$). Hence, for a non-zero $\vec m^{\rm orb} (\vec k)$ ($\vec \Omega (\vec k)$), either of these two symmetries must be broken. In the present case, the broken $\cal I$ symmetry leads to non-zero values of $\vec m^{\rm orb} (\vec k)$ and $\vec \Omega (\vec k)$. We plot our calculated $\vec m^{\rm orb} (\vec k)$ and $\vec \Omega (\vec k)$ in Fig. \ref{fig3}. Note that since $\cal T$ symmetry is preserved, $\vec m^{\rm orb} $ ($\vec \Omega$) at $+\vec k$ has the opposite sign to that at $-\vec k$, and as a result the sum of $\vec m^{\rm orb} (\vec k)$ ($\vec \Omega (\vec k)$) over the occupied part of the Brillouin zone (BZ) is zero, consistent with the overall nonmagnetic behavior of PTO.

 \begin{figure}[b]
\includegraphics[width=\columnwidth]{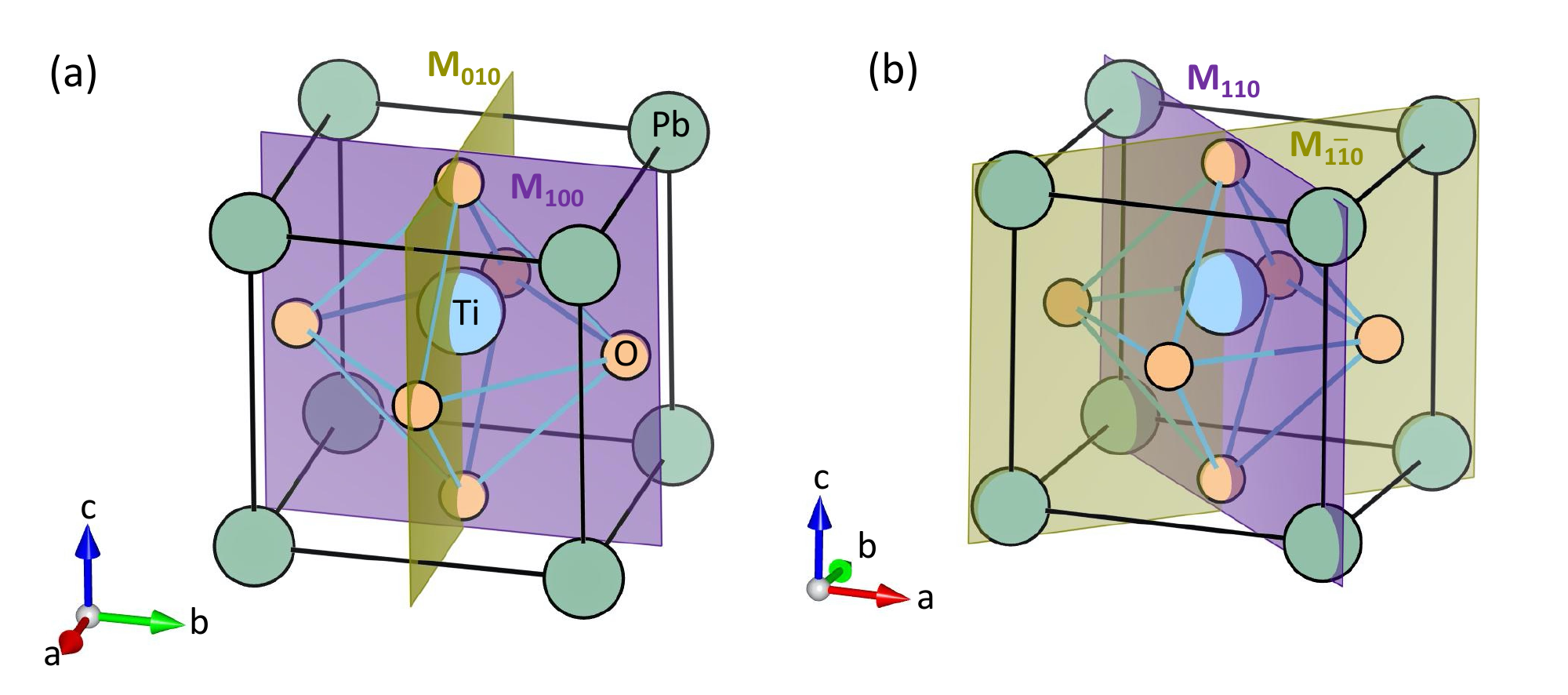}
\caption{(a) Vertical ($\sigma_v$) and (b) diagonal ($\sigma_d$) mirror planes in PTO.}
\label{fig6}
\end{figure}

The key features of the computed distributions in Figs. \ref{fig3} (a)-(d) are the following. First of all, $m^{\rm orb}_x$ ($\Omega_x$) is equal and opposite at $\pm k_y$, while it has the same sign at $\pm k_x$, consistent with the $\sigma_v$ mirror symmetries (see Fig. \ref{fig6} (a)) that dictate: 
\begin{eqnarray}\nonumber
   && m^{\rm orb}_x (k_x,k_y,k_z) \xrightarrow{M_{100}} m^{\rm orb}_x (-k_x,k_y,k_z) \\
\text{and}  && m^{\rm orb}_x (k_x,k_y,k_z) \xrightarrow{M_{010}} -m^{\rm orb}_x (k_x,-k_y,k_z).   
\end{eqnarray} 
 In contrast, $m^{\rm orb}_y$ ($\Omega_y$) is equal and opposite at $\pm k_x$, while having  the same sign at $\pm k_y$ due to the same $\sigma_v$ symmetries, that is,
\begin{eqnarray}\nonumber
   && m^{\rm orb}_y (k_x,k_y,k_z) \xrightarrow{M_{100}} -m^{\rm orb}_y (-k_x,k_y,k_z) \\
\text{and}  && m^{\rm orb}_y (k_x,k_y,k_z) \xrightarrow{M_{010}} m^{\rm orb}_y (k_x,-k_y,k_z).  
\end{eqnarray} 
Furthermore, the $x$ and $y$ components of $\vec m^{\rm orb} (\vec k)$ ($\vec \Omega(\vec k)$) are related to each other by the mirror $M_{1\bar{1}0}$ symmetry (see Fig. \ref{fig6} (b)), viz., $m^{\rm orb}_x (k_x,k_y,k_z)  \xrightarrow{M_{1\bar{1}0}} -m^{\rm orb}_y (k_y,k_x,k_z)$. 
Moreover, since the velocity operator transforms as $(v_x,v_y,v_z)\xrightarrow{M_{1\bar{1}0}}(v_y,v_x,v_z)$ under the mirror $M_{1\bar{1}0}$ symmetry, Eq. (\ref{response1}) [Eq. (\ref{response2})] leads to the constraint $\tilde{\cal K}_{xy}=-\tilde{\cal K}_{yx}$ [${\cal D}_{xy}=-{\cal D}_{yx}$], in agreement with our results in Fig. \ref{fig1} (c) [Fig. \ref{fig1} (d)]. 

\subsection{Microscopic origin: role of odd-parity charge multipoles}\label{TB-analysis}

{\it Model Hamiltonian-} To understand the microscopic origin of these effects, we construct a minimal tight-binding (TB) model in the basis set of the Ti-$t_{2g}$ orbitals, $\{ d_{xy}, d_{yz}, d_{xz}\}$. For small doping, the doped electrons occupy the bands around the $\Gamma$ point of the BZ that correspond to the CBM for the undoped case, indicated by the black circles in Fig. \ref{fig3}. We, therefore, expand the TB model around the $\Gamma$ point, and the resulting low energy model Hamiltonian is given by
\begin{eqnarray}\label{total}
{\cal H}(\vec k) = {\cal H}_{\rm inv}(\vec k) + {\cal H}_{\rm BI}(\vec k). 
\end{eqnarray}
Here ${\cal H}_{\rm inv}$ is the inversion symmetric part of the Hamiltonian, and is given by,
\begin{equation} \label{inv}
	\cal{H}_{\rm inv} = \begin{pmatrix}
		h_{11} & h_{12} & h_{13}\\
		h_{12} & h_{22} & h_{23} \\
		h_{13} & h_{23} & h_{33}
	\end{pmatrix}, 
\end{equation}
with the explicit analytical forms of the elements $h_{ij}$ up to quadratic order in $k$ given below:
\begin{eqnarray} \label{teff} \nonumber
h_{11} &=& t_{\text{eff}}^1 - t_{\text{eff}}^2( k_x^2 + k_y^2)a^2 - t_{\text{eff}}^3 k_z^2 c^2 \\ \nonumber
h_{22} &=& t_{\text{eff}}^4 - t_{\text{eff}}^5 k_x^2 a^2 - t_{\text{eff}}^6 k_y^2 a^2 - t_{\text{eff}}^7k_z^2 c^2 \\ \nonumber
h_{33} &=& t_{\text{eff}}^4 - t_{\text{eff}}^6k_x^2 a^2 - t_{\text{eff}}^5 k_y^2 a^2 - t_{\text{eff}}^7 k_z^2 c^2 \\ \nonumber
h_{12} &=&   t_{\rm eff}^8 k_x k_z ac  \\ \nonumber 
h_{13} &=&  t_{\rm eff}^8 k_y k_z ac  \\ 
h_{23} &=&  t_{\rm eff}^9 k_x k_y a^2. 
\end{eqnarray} 
%
Here $a$ and $c$ are the lattice constants for the tetragonal unit cell. 
Note that since $\cal{H}_{\rm inv}$ is inversion symmetric, it contains only terms that are even in $k$. The effective hopping parameters $t_{\text{eff}}^i$, $i=1,9$ are linear combinations of the different effective $t_{2g}$-$t_{2g}$ electronic hopping parameters that we extract using the N$^{\rm th}$ order muffin-tin orbital (NMTO) downfolding technique \cite{nmto}. The computed parameters for one direction of polarization ($+P$) are listed in Table \ref{tab1}. We considered up to fourth nearest neighbor (NN) interactions. It is important to consider further neighbor interactions which are needed to capture the physics of the two effects of interest, as we discuss later. 

\begin{table} [t]
\caption{Effective hopping parameters (in units of $10^{-2}$ Ry) in Eq. (\ref{teff}), derived from the computed TB hopping parameters and onsite energies for PTO using the NMTO downfolding technique.}
\setlength{\tabcolsep}{4pt}
\centering
\begin{tabular}{ c c  c    c c c c  c c}
\hline
$t^1_{\rm eff}$ & $t^2_{\rm eff}$ & $t^3_{\rm eff}$ & $t^4_{\rm eff}$ & $t^5_{\rm eff}$ & $t^6_{\rm eff}$ & $t^7_{\rm eff}$ & $t^8_{\rm eff}$ & $t^9_{\rm eff}$ \\ [1 ex]
\hline\hline
 4.95 & -2.26 & 0.4 & 10.09  & -0.19 & -0.97 & -1.59 & -0.48 & -0.28 \\
  \hline
\end{tabular}
\label{tab1}
\end{table}

On the other hand, ${\cal H}_{\rm BI}$ includes the hopping parameters that are induced by the broken ${\cal I}$ symmetry. It can be written in terms of the components of the orbital angular momentum operator $\hat{\vec L}$, 
\begin{eqnarray} \nonumber \label{tbm}
    {\cal H}_{\rm BI} &=& \frac{\alpha a}{\hbar}(k_x\hat{L}_y-k_y\hat{L}_x) -\frac{\alpha a^3}{6\hbar}(k_x^3\hat{L}_y-k_y^3\hat{L}_x) \\ \nonumber
    &-& \frac{\beta ac^2}{\hbar}k_z^2(k_x\hat{L}_y-k_y\hat{L}_x)-\frac{\gamma a^3}{\hbar}k_xk_y (k_y\hat{L}_y-k_x\hat{L}_x). \\
\end{eqnarray}
The parameters $\alpha,\beta, \gamma$ are determined by the broken ${\cal I}$-symmetry-induced hopping parameters and have opposite signs for $+P$ and $-P$. In centrosymmetric PTO, $\alpha,\beta, \gamma$ are zero so that  $\cal{H}=\cal{H}_{\rm inv}$. In addition, $t_{\rm eff}^8=-2( t^x - t^y)=0$ in the centrosymmetric structure, where $t^x$ and $t^y$ are the fourth nearest neighbor inter-orbital hopping integrals, which we discuss in detail later.

The components of the orbital angular momentum operator in Eq. (\ref{tbm}) in the $t_{2g}$ orbital basis $\{ d_{xy}, d_{yz}, d_{xz}\}$ are given by, 
\begin{eqnarray}\label{Lop} \nonumber
	 L_x^{(t_{2g})} =&&\hbar\begin{pmatrix}
		0 & 0 & -i \\
		0 & 0 & 0 \\
		i & 0 & 0
	\end{pmatrix},  \quad L_y^{(t_{2g})} = \hbar\begin{pmatrix}
		0 & i & 0 \\
		-i & 0 & 0\\
		0 & 0 & 0
	\end{pmatrix} , \\ \newline 
  && L_z^{(t_{2g})} = \hbar\begin{pmatrix}
 		0 & 0 & 0	\\
 		0 & 0 & i \\
 		0 & -i & 0	
		\end{pmatrix} .
\end{eqnarray} 

\begin{figure}[t]
\includegraphics[width=\columnwidth]{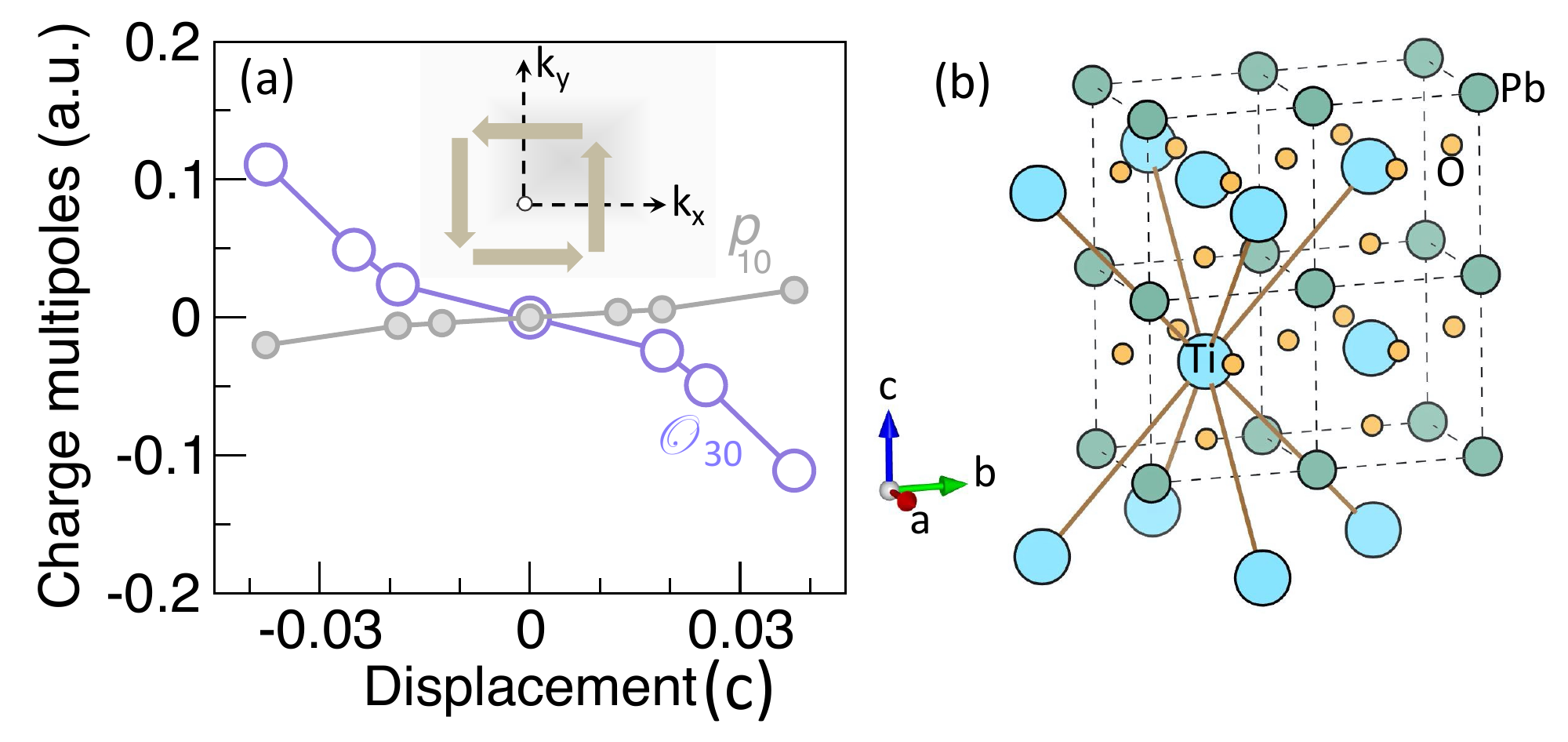}
\caption{Atomic-site charge dipole moment component $p_{10}$ and octupole moment component ${\cal O}_{30}$ on the Ti$^{4+}$ ions as a function of the displacement (in units of out-of-plane lattice constant $c$) of the Ti ion from the center of the unit cell in PTO. The inset shows the schematic for the toroidal arrangement of the orbital (spin) moment (indicated in thick arrows) in the $k_x$-$k_y$ plane due to the first term in the Hamiltonian (\ref{tbm}) driven by the charge dipole. (b) Fourth nearest neighbor Ti atoms (connected by the brown straight lines) along $(\pm a, 0, \pm c)$ and $(0, \pm a, \pm c)$. Note that in the cubic high-symmetry structure with $c=a$, these are second nearest neighbors. }
\label{fig4}
\end{figure}

The advantage of writing ${\cal H}_{\rm BI}$ in terms of the $\hat{\vec L}$ operators is that we can readily identify the resulting orbital texture in momentum space. For example, the first term in Eq. (\ref{tbm}), which is linear in $\vec k$, depicts a toroidal arrangement of orbital magnetic moment in reciprocal space (see the inset of Fig. \ref{fig4} (a)). Such a toroidal arrangement of the orbital moment in $k$ space is also in agreement with our DFT results (see Fig. \ref{fig3} (a) and (b)) and the symmetry analysis presented in section \ref{distribution}. We note that the first term in Eq. (\ref{tbm}) has a form $\sim (\vec k \times \vec L)$, which is an orbital counterpart of the (spin) Rashba effect $\sim (\vec k \times \vec \sigma)$ and, hence, is often referred to as an orbital Rashba effect \cite{Go2017,GoPRB2021}. In the presence of SOC, the orbital texture in the orbital Rashba effect couples to the spin, additionally leading to spin texture and the Rashba effect in PTO \cite{Arras2019}. The Rashba spin-splitting $\Delta \varepsilon_s (\vec k)$ is antisymmetric in $\vec k$, corresponding to $p$-wave symmetry, due to the presence of time-reversal symmetry, which means that $\Delta \varepsilon_s (\vec k) = \varepsilon_{\uparrow} (\vec k)- \varepsilon_{\downarrow} (\vec k) = -\Delta \varepsilon_s (-\vec k)$. Here, for simplicity, we do not include SOC in our model Hamiltonian in Eq. (\ref{total}), since both KME and BCD exist even in its absence (See Fig. \ref{fig2}).   

{\it Role of odd-parity charge multipoles-} Interestingly, each term of different order in $\vec k$ in the Hamiltonian ${\cal H}_{\rm BI}$ of Eq. (\ref{tbm}) has a direct correlation to a corresponding odd-parity charge multipole. Recently, we showed that the $k$-space orbital and spin textures in ferroelectrics result from the $k$-space magnetoelectric multipoles that are reciprocal to the real-space odd-parity charge multipoles \cite{BhowalCollinsSpaldin2022}. The odd-parity charge multipoles characterize the asymmetries in the charge density that are present due to the broken ${\cal I}$ symmetry. For example, the electric dipole dictates the first-order asymmetry in the charge density, while the electric octupole corresponds to the third-order asymmetry, and so on. The first term within the parentheses in Eq. (\ref{tbm}), which is linear in $\vec k$, corresponds to the $k$-space representation of the electric dipole moment ($p_{10}$) whereas the remaining terms, which are all cubic in $\vec k$, correspond to the electric octupole moment (${\cal O}_{30}$). 

To verify the existence of the local electric dipoles and octupoles in PTO, we decompose the $\cal T$ symmetric density matrix $\rho_{lm,l'm'}$, computed within the DFT framework, into parity-odd tensor moments and explicitly compute the atomic-site electric dipole and octupole moments, for which only the odd $l-l'$ terms contribute \cite{Spaldin2013}. The computed odd-parity charge multipoles on the Ti${^4+}$ ions are non-zero in the polar structure, as shown in Fig. \ref{fig4} (a), and confirm the presence of a ferrotype ordering of electric dipole component $p_{10}$ and octupole component ${\cal O}_{30}$ at the Ti site. Here the indices at the suffix of the multipole components represent the $l$ and $m$ indices of the spherical harmonics that are used to build these charge multipoles. The electric dipole moment $\vec p$ is a tensor of rank 1 (vector), with $p_{10}$ indicating its $z$ component. Similarly, the octupole moment ${\cal O}_{ijk}$ is a totally symmetric tensor of rank 3 with seven components. The ${\cal O}_{30}$ component has the representation $\frac{1}{2}z(5z^2-r^2)$. 

{\it Results and discussion-} Now that we have correlated the individual terms of the Hamiltonian to the charge multipoles, we diagonalize the Hamiltonian ${\cal H}(\vec k)$ in Eq. (\ref{total}) for the realistic parameters listed in Table \ref{tab1}, extracted using the NMTO downfolding technique \cite{Lowdin2004,AndersonSahaDasgupta2000}.  
We, then, use the computed eigenvalues $\epsilon_k^n$ and eigenfunctions $u_k^n$ to obtain the $k$-space distribution of the orbital moment and the Berry curvature using Eqs. (\ref{morb}) and (\ref{BC}) for the lowest energy band of the Hamiltonian in Eq. (\ref{total}). 

\begin{figure}[t]
\includegraphics[width=\columnwidth]{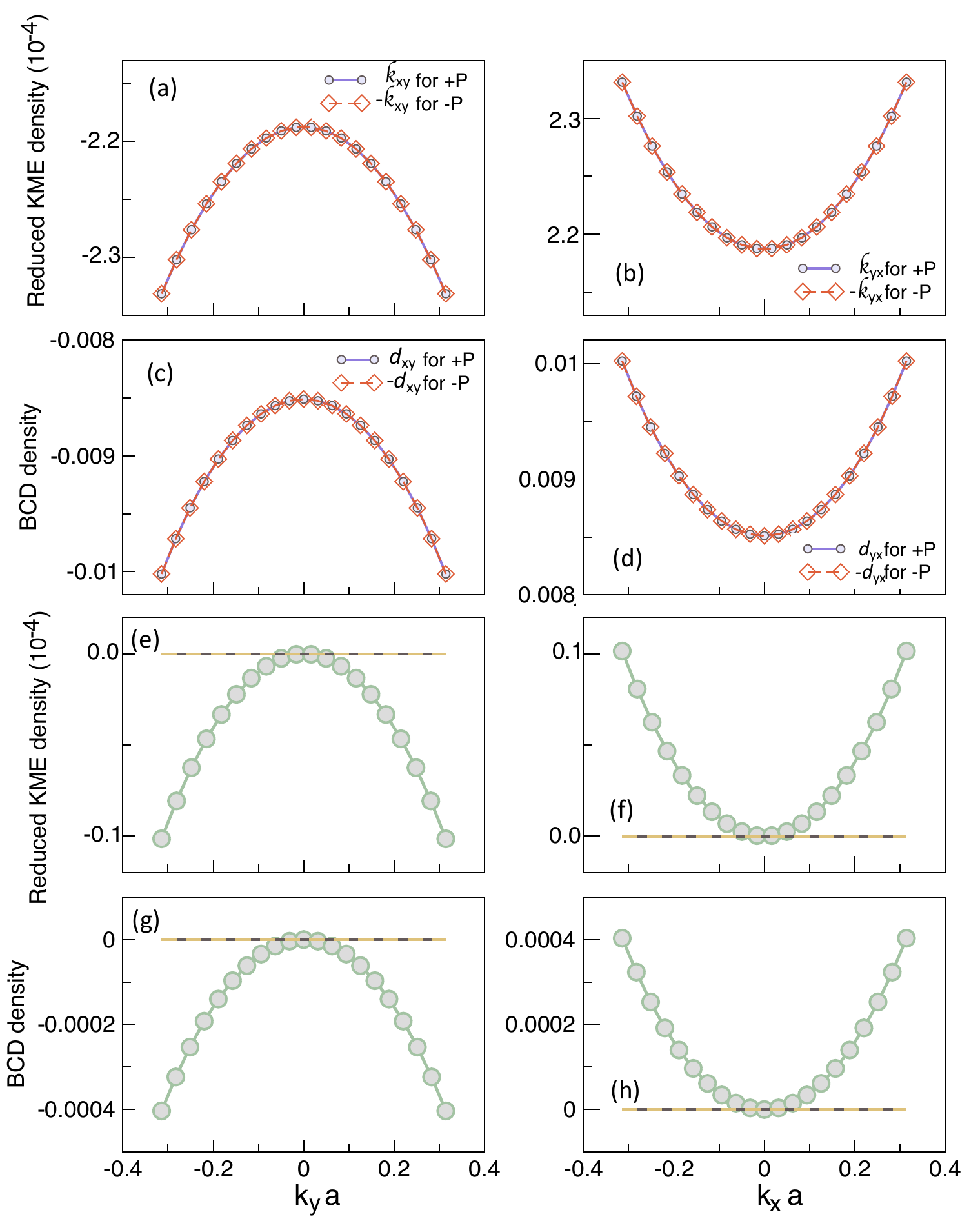}
\caption{Results of the tight-binding analysis. Computed variation (solid line with circles) of the reduced KME density components (a) $\kappa_{xy}$, (b) $\kappa_{yx}$, and the BCD density components (c) $d_{xy}$, (d) $d_{yx}$ around the $\Gamma$ point for the $+P$ polarization. The variation is shown along $k_y$ for (a) and (c), and along $k_x$ for (b) and (d). The same variation (indicated in dashed lines with diamonds) for $-\kappa_{xy}$, $-\kappa_{yx}$, $-d_{xy}$, and $-d_{yx}$ for the polarization $-P$ are also shown in (a)-(d). 
The same variation of (e) $\kappa_{xy}$, (f) $\kappa_{yx}$, (g) $d_{xy}$ and (h) $d_{yx}$ in the presence of inversion symmetry (black solid line), in absence of fourth NN inter-orbital hopping parameters $t^x$ and $t^y$ (dashed brown line), and in absence of the first term (linear in $\vec k$) in Eq. (\ref{tbm}) (green line with circles).  
The parameters used for the plots are listed in Table \ref{tab1}, and $\alpha=0.22, \beta=0.02,$ and $\gamma=-0.10$ (in units of 10$^{-2}$ Ry) for $+P$ polarization.}
\label{fig5}
\end{figure}

Note that the second term in Eq. (\ref{morb}) does not contribute to the KME response, as stated before, and hence, we ignore this term for the computation of the orbital moment.   
We then compute the BCD density $d_{ij}(\vec k) = \partial_{k_i} \Omega_j (\vec k)$ and the reduced KME density ${\kappa}_{ij} (\vec k) = \partial_{k_i} m^{\rm orb}_j (\vec k)$ for $i,j=x,y$, the integrals of which over the occupied part of the BZ determine the magnitude of ${\cal D}_{ij}$ and $\tilde{\cal K}_{ij}$ respectively [see Eqs. (\ref{response1}) and (\ref{response2})]. The computed densities show that they have the same sign (+ or -) over $k$-space only if $i \ne j$ and hence when integrated over the occupied part of the BZ, only the $xy$ and $yx$ components of ${\cal D}$ and $\tilde{\cal K}$ have non-zero values. The variations of these components along a specific momentum direction are shown in Fig. \ref{fig5} (see the solid lines). 

For the opposite polarization direction ($-P$), the parameters $\alpha, \beta, \gamma$ switch sign and, consequently, as shown in Fig. \ref{fig5}, the $xy$ and $yx$ components of $d$ and $\kappa$ switch signs, keeping their magnitudes unaltered. In an ${\cal I}$-symmetric system, on the other hand, $\alpha=\beta=\gamma=0$, and consequently, we find that $d_{ij}$, $\kappa_{ij}$ become zero as shown in the insets of Fig. \ref{fig5}, emphasizing the important role of ${\cal I}$ symmetry breaking. 

Further to gain insight into the origin of these two effects, we switch off the linear term in Eq. (\ref{tbm}), which originates from the electric dipole moment. Interestingly, in this case, we find that while all the considered components of $d$, $\kappa$ still survive, their values reduce drastically by an order of magnitude. This suggests that the linear terms in $k$ in Eq. (\ref{tbm}), originating from the electric dipole moment, play an important role in determining the magnitudes of both these effects, although the importance of the electric octupole-driven $k^3$ terms can not be ignored. Our findings are consistent with the multipole description of the KME response, proposed by Hayami {\it et. al.} based on symmetry analysis \cite{Hayami2018}. Indeed, we find that the antisymmetric part of the KME response ${\cal K}^{-}_{ij}$ in PTO can be described by the existence of an  electric dipole moment component, $\tilde{\cal K}^-_{ij}= \frac{1}{2}({\cal K}_{ij}-{\cal K}_{ji})=\varepsilon_{ijk} p_k$. It is important to point out here that the KME, although universal to all polar metals, can also occur in noncentrosymmetric but non-polar systems, e.g., chiral materials, in which case  other multipoles such as the monopole of the electric toroidal dipole moment will dictate the symmetric part (with the trace) of the KME response \cite{Hayami2018}.

We further note that the fourth NN (see  Fig. \ref{fig4} (b)), inter-orbital ($d_{xy}-d_{xz}$ and $d_{xy}-d_{yz}$) hopping integrals, $t^x$ and $t^y$, induced by the broken $\cal I$ symmetry, are the key ingredients for both these effects. While both these hopping integrals contribute to the parameters $\alpha$ and $\beta$, $\beta$ is solely determined by $t^x$ and $t^y$ while  $\alpha$ has additional contributions. As a result, in the absence of these hoppings, $\beta$ and the effective hopping, $t_{\rm eff}^8$, in $\cal{H}_{\rm inv}$ vanish. In this case of $t^x=t^y=0$, we find that the components of both $d$ and $\kappa$ also vanish, as shown in the insets of Fig. \ref{fig5} (see the dashed brown line), emphasizing the importance of the further neighbor interactions.

To understand why the fourth NN hopping parameters are crucial, we first note that the non-zero $\beta$ and $t_{\rm eff}^8$ resulting from the fourth NN hopping parameters appear in the third term of Eq. (\ref{tbm}) and the off-diagonal elements $h_{12}$ and $h_{13}$ of Eq. (\ref{inv}) respectively. Interestingly, these are the only inter-orbital contributions in our minimal model that are also responsible for the band dispersion along the out-of-plane $k_z$ direction. Since the inter-orbital hopping parameters drive the non-zero Berry curvature \cite{BhowalSatpathy2019} and since the dispersion along $k_z$ is crucial for the existence of the in-plane components of both orbital moment and Berry curvature [see Eqs. (\ref{morb}) and (\ref{BC})], we see that both quantities vanish in the absence of fourth NN hopping. This, in turn, also leads to an absence of $xy$ and $yx$ components of $d$ and $\kappa$, explaining the crucial role of the fourth NN hopping integrals in driving the KME and NHE in doped PTO.

\section{Summary and Outlook} \label{summary}

To summarize, taking the example of doped PTO, we have shown that both the KME and the NHE, are universal to all polar metals and can be used for a complete characterization of this class of materials. Our work paves the way for the broad applicability of these two effects in polar metals in general, going beyond their earlier investigation in topological systems \cite{Zhong2016,Johansson2018,Tsirkin2018,Roy2022}. Our detailed tight-binding analysis reveals the importance of the broken-symmetry-induced inter-orbital hopping parameters, correlated to the odd-parity charge multipoles, in mediating these effects. In particular, we have identified the broken-inversion-induced fourth NN inter-orbital hopping parameters as being essential in driving these effects in doped PTO.

{\it Proposal for experiments.} Before concluding, here we briefly discuss possible routes to detecting the two effects. The second-order NHE in polar metals can be detected by measuring the second harmonic current $J_{2\omega}$ at a frequency $2\omega$ for an applied ac electric field $\vec E$ of frequency $\omega$, \cite{SodemannFu2015} 
\begin{equation} \label{exp}
 \vec j_{2\omega}= \frac{e^3\tau}{2(1+i\omega\tau)} {\vec E_{\omega}} \times ({\vec p} \times {\vec E_{\omega}}).   
\end{equation}
Here $\vec p$ is the direction of the electric dipole moment, which is along $\hat z$ for doped PTO. This suggests that for $\vec E$ along $\hat z$ (i.e., with polar angle $\theta=0$), the Hall current vanishes as we found also from our explicit calculations discussed above. Furthermore, for a general form of the field, $\vec E = E e^{i\omega t}(\sin \theta \cos \phi, \sin \theta \sin \phi, \cos \theta)$, it is also easy to see from Eq. (\ref{exp}) that the Hall current does not depend on the azimuthal angle $\phi$ made by $\vec E$ with $\hat x$ for an in-plane $\vec E$ (i.e., $\theta=\pi/2$).
This means that rotation of $\vec E$ within the $x$-$y$ plane will leave the Hall current invariant. 

The current-induced magnetization in the KME should be detectable using the magneto-optical Kerr effect. 
In doped PTO the generated magnetization is dominated by the orbital moment for a reasonable doping concentration (see the inset of Fig. \ref{fig1} (c)) and has a magnitude of $~1.8 \times 10^{-4} \mu_B/$atom at the experimentally observed maximum doping concentration ($n_{x=0.12}=1.9 \times 10^{21}$ cm$^{-3}$) up to which the system retains the ferroelectricity, for an applied field of $10^5$ V/m and a typical relaxation time constant $\tau \simeq 1$ ps. The computed orbital magnetization is about four orders of magnitude larger than that reported in Te \cite{Tsirkin2018}, while it is about an order of magnitude smaller than the orbital magnetization in BCC iron \cite{Lopez2012}. The computed total (spin plus orbital) magnetization of $ \sim 1.0\times10^{-3} \mu_B$ per unit cell is also comparable to the magnetization of the Rashba system Bi/Ag(111), the (001) surface of the topological insulator $\alpha$-Sn, and the Weyl semimetal TaAs \cite{Johansson2018} and, hence, likely to be discernible in measurements.

In the present work, we considered a rigid band approximation to describe the doped PTO case. While we expect this to provide a good description of the NHE and KME for the small doping concentration achievable in the measurements, future work should investigate computationally how electron doping affects the electronic structure of PTO. The dominance of the orbital magnetization in the KME response of doped PTO that emerges from our work, opens the door for the application of polar metals in orbitronics with the additional advantage of switchable orbital texture by reversal of the electric polarization. We hope that our work will motivate both theoretical and experimental work in these directions in the near future.

\section*{Acknowledgements}
The authors thank Awadhesh Narayan and Dominic Varghese for stimulating discussions.
NAS and SB were supported by the ERC under the EU’s Horizon 2020 Research and Innovation Programme grant No 810451 and by the ETH Zurich. Computational resources were provided by ETH Zurich's Euler cluster, and the Swiss National Supercomputing Centre, project ID eth3.

\bibliography{Reference}

\begin{thebibliography}{47}
\bibitem[{gyr()}]{gyrotropic}
\bibinfo{note}{The gyrotropic point groups are those crystal point groups that
  allow the presence of both polar ($\vec P$) and axial ($\vec L$) vector
  components transforming according to equivalent representations of the point
  group. The 18 gyrotropic point groups, therefore, allow for a second-rank
  tensor, connecting physical quantities $P_i$ and $L_j$, that describes
  physical effects such as natural optical activity.}
\end{thebibliography}

\end{document}